\documentclass[osajnl,twocolumn,showpacs,superscriptaddress,10pt]{revtex4-1} 
\usepackage{amsmath,amssymb,graphicx}
\usepackage{epstopdf}

\begin{document}

\title{Aberrations in shift-invariant linear optical imaging systems using partially coherent fields}

\author{Mario A. Beltran}\email{Corresponding author: Mario.Beltran@monash.edu}
\affiliation{School of Physics, Monash University, Victoria 3800,
Australia}

\author{Marcus J. Kitchen}
\affiliation{School of Physics, Monash University, Victoria 3800,
Australia}

\author{T. Petersen}
\affiliation{School of Physics, Monash University, Victoria 3800,
Australia}

\author{David M. Paganin}
\affiliation{School of Physics, Monash University, Victoria 3800,
Australia}

\begin{abstract}
Here the role and influence of aberrations in optical imaging systems employing partially coherent complex scalar fields is studied. Imaging systems require aberrations to yield contrast in the output image. For linear shift--invariant optical systems, we develop an expression for the output cross--spectral density under the space--frequency formulation of statistically stationary partially coherent fields. We also develop expressions for the output cross--spectral density and associated spectral density for weak--phase, weak--phase--amplitude, and single--material objects in one transverse spatial dimension.
\end{abstract}




\maketitle 

\section{Introduction} 

When imaging transparent samples in an in-focus optical system such as a visible--light or x--ray microscope, the detected output image appears almost featureless if the system yields a reproduction of the input image that is incident upon the system \cite{Zernike01}. This is what in optics is commonly defined as a perfect or near perfect imaging system in which there are no transverse spatial variations within the incident spectral density distribution as it propagates to the output detection plane. Note that the term ``spectral density'' is here used in the sense of optical partial coherence. As perfect systems are unable to visualize the refraction effects (phase contrast) caused by transparent samples, the presence of aberrations is a necessary condition for non--negligible contrast in the output spectral density to be attained \cite{Pag01}. In this context, an aberrated imaging system may be defined as one whose output transverse spatial distribution of spectral density is not equal to the input transverse spatial distribution of spectral density, up to transverse and multiplicative scale factors together with the smearing effects of finite resolution. Almost all aberrated imaging systems exhibit phase contrast, i.e. have an output spatial distribution of spectral density which is influenced by the functional form of the input wavefronts (input phase distribution). Examples of aberrated imaging systems yielding phase contrast include Zernike phase contrast, propagation--based phase contrast, differential phase contrast, inline holography, etc. \citep{Zernike01,Wilkins01,Gab01,Forster01}

Work relating to a partially coherent treatment specifically for propagation--based phase contrast imaging based on the Transport--of--Intensity equation has been reported \citep{Zusk01,Jon01,Gureyev01}. In this paper we consider the generalized differential phase contrast associated with aberrated linear shift--invariant optical imaging systems employing statistically stationary partially coherent scalar radiation, for which the output spatial distribution of spectral density (i.e., the output image) can be modelled using the transfer function formalism. This extends previously reported work by Paganin and Gureyev \cite{Pag01} which restricted consideration to the generalized differential phase contrast of fully coherent scalar fields imaged using aberrated linear shift--invariant optical systems.

In Sec.~\ref{section1.2D} we obtain an equation that describes the action of shift--invariant linear systems using partially coherent fields, under the imaging assumption that the object under study is a pure thin phase object. A two--dimensional transverse Cartesian coordinate system is used in the derivation. In Sec.~\ref{1Dsect} expressions for the spectral density are derived, restricting consideration to only one transverse spatial variable for simplicity. Three different types of sample are considered: Samples that satisfy, i) the weak--phase object approximation, ii) the weak phase--amplitude approximation and; iii) the single material weak phase--amplitude approximation. Sec.~\ref{TransSec} studies in depth the features of the transfer function used in this formalism.


\section{Shift--invariant, linear systems for partially coherent fields using two transverse spatial coordinates} \label{section1.2D}

In this section we derive an expression for partially coherent complex scalar fields imaged by an optical system that is shift--invariant and satisfies the property of linearity \cite{Good01}. For such a system, the output complex disturbance is related to the input complex disturbance by the transfer function formalism \cite{Good01}. Since most image collecting is normally done using two dimensional Cartesian grids it is natural to utilize a two--dimensional Cartesian system $(x,y)$ in all calculations.

Before incorporating the effects of partial coherence in our derivations, we recall first a description of shift--invariant linear systems for fully coherent complex scalar wave--fields which are governed by the transfer function formalism. For such optical systems the output field $\Psi_{out}(x,y)$ is related to the input field $\Psi_{in}(x,y)$ by a Fourier--space filtration that can be written in operator form as \cite{Pag01}:

\begin{eqnarray} \label{Oper1}
\Psi_{out}(x,y)=F^{-1}\overline{T}(k_{x},k_{y})F\left \{ \Psi_{in}(x,y)
\right \}.
\end{eqnarray}

\begin{figure}[ht!]
\includegraphics[scale=0.9]{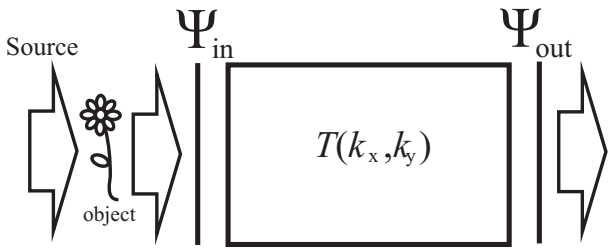}
\caption{Schematic illustration of the action of an aberrated shift--invariant linear optical system for imaging fully coherent complex scalar wave--fields, under the transfer function formalism. Input and output complex fields are related by the transfer function formalism according to Eq.\eqref{Oper1}.} 
\label{fig1}
\end{figure}

\noindent Here, $\overline{T}(k_{x},k_{y})$ is the transfer function characterizing the optical system, $(k_{x},k_{y})$ are Fourier conjugate coordinates dual to $(x,y)$, $F$ and $F^{-1}$ respectively represent the forward and inverse Fourier transform operations, and all operators are taken to act from right to left. Thus, the above equation states that $F$ is applied to the input field $\Psi_{in}(x,y)$, before multiplying by the transfer function $\overline{T}(k_{x},k_{y})$ and then applying the operator $F^{-1}$, so as to yield the output field $\Psi_{out}(x,y)$ (see Fig.~\ref{fig1}).

In our derivation the forward and inverse Fourier transform operation conventions used are the following:

\begin{subequations}
\begin{align}
\widehat{G}(k_{x},k_{y})=\frac{1}{2\pi}\iint_{-\infty}^{\infty}dxdy\;G(x,y)e^{ -i(k_{x}x+k_{y}y)}  \label{Conven1:a}, \\
G(x,y)=\frac{1}{2\pi}\iint_{-\infty}^{\infty}dk_{x}dk_{y}
\widehat{G}(k_{x},k_{y})e^{ i(k_{x}x+k_{y}y)}. \label{Conven1:b}
\end{align}
\end{subequations}

\noindent Here, $\widehat{G}(k_{x},k_{y})\equiv F\left
\{ G(x,y) \right \}$.

To proceed further, we follow Paganin and Gureyev \cite{Pag01} and make the restricting assumption that the transfer function $\overline{T}(k_{x},k_{y})$ is sufficiently well behaved for its logarithm to admit a Taylor--series representation. Note that a necessary condition for this assumption to be valid is that the transfer function does not possess any zeros over the patch of Fourier space for which the modulus of $F\left \{\Psi_{in}(x,y)  \right \}$ is non--negligible, a region which may be termed the ``essential spectral support'' of the input field.

While this key assumption will fail for imaging systems such as Schlieren optics which completely block certain spatial frequencies in the essential spectral support of the input disturbance, the assumption will hold for a variety of important imaging systems such as out--of--focus contrast \cite{Wilkins01}, inline holography \cite{Gab01}, interferometric phase contrast \cite{Bonse01}, differential phase contrast \cite{Pfeiffer01}, and analyzer--based phase contrast of weakly scattering samples \cite{Forster01} etc.

With the above in mind, our simplifying assumption allows us to
express the transfer function in the classic form that is standard e.g. in transmission electron microscopy, namely \citep{Pag01,Pag03,Cow01}:

\begin{eqnarray} \label{Transf1}
\overline{T}(k_{x},k_{y})=\exp\left ( i\sum_{m,n=0}^{\infty }\widetilde{\alpha}_{mn}k_{x}^{m}k_{y}^{n}  \right ).
\end{eqnarray}

\noindent Under this representation we denote the set of complex numbers $\left \{ \widetilde{\alpha}_{mn} \right \}$ as the ``aberration coefficients'' where $m$ and $n$ are non--negative integers and label the order of the aberration.  The real part of each such coefficient is termed a coherent aberration, with the corresponding imaginary part being termed an incoherent aberration.  See Paganin and Gureyev \cite{Pag01} for a direct link between these complex aberration coefficients, and the Siedel aberrations \cite{Wolf02} (e.g., piston, defocus, astigmatism, spherical aberration, chromatic aberration etc.) of classical aberration theory.

Expanding the complex exponential in Eq.~\eqref{Transf1} as a
Taylor--series, we obtain:

\begin{eqnarray} \label{Transf2}
\overline{T}(k_{x},k_{y})=1+i\sum_{m,n=0}^{\infty }\alpha_{mn}k_{x}^{m}k_{y}^{n}.
\end{eqnarray}

The above expression serves to define the set of coefficients $\left \{ \alpha_{mn} \right \}$ . The set of coefficients $\left \{ \alpha_{mn} \right \}$ is defined in terms of the set of aberration coefficients $\left \{ \widetilde{\alpha}_{mn} \right\}$. We note that like Eq.~\eqref{Transf1}, Eq.~\eqref{Transf2} disallows the presence of any zeros in the transfer function $\overline{T}(k_x,k_y)$. This form is particularly useful for studying the effect of transfer functions which differ only slightly from unity, namely for weakly aberrated shift--invariant imaging systems. We shall pick up on this point later in the paper.

It is useful to write the operator form of Eq.~\eqref{Oper1} in terms of the following integral:

\begin{eqnarray}
\Psi_{out}\left ( x,y \right )&=&\frac{1}{2\pi}\iint_{-\infty}^{\infty}dk_{x}dk_{y}
\overline{T}(k_{x},k_{y})e^{ i(k_{x}x+k_{y}y)} \nonumber\\
& & \times \widehat{\Psi}_{in}(k_{x},k_{y}), \nonumber\\
\end{eqnarray}

\noindent where $\widehat{\Psi}_{in}(k_{x},k_{y})$ denotes the Fourier transform of $\Psi_{in}(x,y)$ with respect to $x$ and $y$. The above integral--form expression describes the output wave--field for an optical system that is linear and shift--invariant for incoming wave--fields that are fully coherent. 

\begin{figure}[ht!]
\includegraphics[scale=1.0]{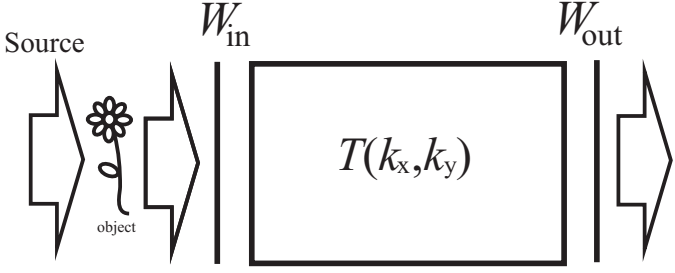} 
\caption{Schematic illustration of the action of an aberrated shift--invariant linear optical imaging system, for statistically stationary partially coherent complex scalar fields, under the transfer function formalism. Input and output cross--spectral densities, $W_{in}$ and $W_{out}$ respectively, are related by the generalized transfer function formalism according to Eq.~\eqref{Cross2}.}
\label{Fig2}
\end{figure}

We now turn to the extension of this theory of fully coherent fields to partially coherent fields. This corresponds to the generalization shown in Fig.~\ref{Fig2}. Here, $W_{in}$ is the cross--spectral density incident upon a linear shift--invariant aberrated optical system, yielding the corresponding output cross--spectral density $W_{out}$.    

Under the space--frequency description of partial coherence developed by Wolf \cite{Wolf02,Wolf01}, the output cross--spectral density at a specified angular frequency $\omega$ may be constructed using an ensemble of strictly monochromatic fields all of the same angular frequency, via:

\begin{eqnarray} \label{Cross1}
W_{out}\left ( x_{1},y_{1},x_{2},y_{2} \right )=\left \langle
\Psi_{out}^{*}(x_{1},y_{1}) \Psi_{out}(x_{2},y_{2}) \right
\rangle_{\omega}.
\end{eqnarray}

\noindent Here, angular brackets denote the ensemble average. Note that one may also consider expressing $W_{out}$ in terms of its coherent mode expansion but incorporating other such correlating descriptions into our framework is beyond the scope of this paper \cite{Wolf02}.   

Putting this equation to one side for the moment, note that we can express $\Psi_{out}^{*}(x_{1},y_{1})$ and $\Psi_{out}(x_{2},y_{2})$ in terms of the Fourier transform of $\Psi_{in}^{*}(x_{1},y_{1})$ and $\Psi_{in}(x_{2},y_{2})$ respectively using the conventions in Eq.~\eqref{Conven1:a} and \eqref{Conven1:b} which give the following:

\begin{subequations} 
\begin{align}
\Psi^{*}_{out}\left ( x_{1},y_{1} \right )&=&\frac{1}{(2\pi)}\iint_{-\infty}^{\infty}dk_{x_{1}}dk_{y_{1}}
\overline{T}^{*}\left ( k_{x_{1}}, k_{y_{1}} \right )\nonumber\\
& & \times e^{ i^{*}(k_{x_{1}}x_{1}+k_{y_{1}}y_{1})}\widehat{\Psi}^{*}_{in}(k_{x_{1}},k_{y_{1}}), \label{Conven2:a} \\
\Psi_{out}\left ( x_{2},y_{2} \right )&=&\frac{1}{(2\pi)}\iint_{-\infty}^{\infty}dk_{x_{2}}dk_{y_{2}}
\overline{T}\left ( k_{x_{2}}, k_{y_{2}} \right )\nonumber\\
& & \times e^{ i(k_{x_{2}}x_{2}+k_{y_{2}}y_{2})}\widehat{\Psi}_{in}(k_{x_{2}},k_{y_{2}}).\label{Conven2:b}  \\  \nonumber
\end{align} \label{Conven2}
\end{subequations}

By substituting the above expressions into Eq.~\eqref{Cross1}, one can obtain the cross--spectral density in terms of the input cross--spectral density as \cite{Gbur01}:

\begin{eqnarray} \label{Cross2}
W_{out}&=&\frac{1}{(2\pi)^{2}}\iiiint_{-\infty}^{\infty}dk_{x_{1}}dk_{y_{1}}dk_{x_{2}}dk_{y_{2}}
\overline{T}^{*}\left ( k_{x_{1}},k_{y_{1}} \right ) \nonumber\\
& & \times \overline{T}\left ( k_{x_{2}},k_{y_{2}} \right ) e^{[i^{*}(k_{x_{1}}x_{1}+k_{y_{1}}y_{1})+i(k_{x_{2}}x_{2}+k_{y_{2}}y_{2})]} \nonumber\\
& & \times \left \langle \widehat{\Psi}^{*}_{in}( k_{x_{1}},k_{y_{1}}) \widehat{\Psi}_{in}( k_{x_{2}},k_{y_{2}}) \right \rangle_{\omega}. \nonumber\\
\end{eqnarray}

Eq.~\eqref{Cross2} can be visualized pictorially in the diagram shown in Fig.~\ref{Fig2}. This three step process, whereby each member of the ensemble with its corresponding weighting factor $\eta ^{j}$ undergoes forward scattering modelled by the appropriate transfer function at each stage, can be viewed as the following operation: 

\begin{eqnarray} \label{3StepsProcess}
\left \{ \Psi_{S}^{j}, \eta ^{j}  \right \}\rightarrow \left \{ \Psi_{in}^{j}, \eta ^{j}  \right \}\rightarrow \left \{ \Psi_{out}^{j}, \eta ^{j}  \right \}.
\end{eqnarray}

\noindent Here, $\Psi_{S}^{j}$ denote the ensemble members corresponding to the source (Before reaching the object). $\Psi_{in}^{j}$ denote the ensemble members of the field once it has traversed the object and $\Psi_{out}^{j}$ denotes the members of the ensemble after going through the imaging system. It is important to note that Eqs.~\eqref{Cross2} and \eqref{3StepsProcess} apply for any elastic scattering induced by the object to produce $W_{in}$, which is then propagated through the generalized imaging system to yield $W_{out}$. Also, in this paper no assumption regarding the particular nature of the statistical ensemble of fields is made beyond the standard ergodicity and stationarity for the mutual coherence function \cite{Wolf01}. In the present, we consider input wave--fields that are described by the ``phase object approximation" which by definition are wave--fields that only vary in phase but not in amplitude, that is $\Psi_{in}(x,y)=e^{i\phi_{in}(x,y)}$, this way our final expression will be a series of terms which have operations on the input phase $\phi_{in}(x,y)$ which is a real function. Later in the paper we generalize to include absorption. Expanding the complex exponential in this expression as a Taylor series, which implies no loss of generality on account of the infinite radius of convergence of this series, we obtain:

\begin{eqnarray} \label{PsiTay2D}
\Psi_{in}( x,y)=1+\sum_{p=1}^{\infty }\frac{i^{p}}{p!}\phi_{in}^{p}( x,y).
\end{eqnarray}

Taking the Fourier transform of the above expression with respect to $x$ and $y$, we can then write down the following expressions for the terms $\widehat{\Psi}^{*}_{in}( k_{x_{1}},k_{y_{1}})$ and $\widehat{\Psi}_{in}( k_{x_{2}},k_{y_{2}})$ in Eq.~\eqref{Cross2}:

\begin{subequations}
\begin{align}
\widehat{\Psi}_{in}^{*}( k_{x_{1}},k_{y_{1}})=\delta(k_{x_{1}},k_{y_{1}})+\sum_{p=1}^{\infty }\frac{(i^{p})^{*}}{p!}\widehat{\phi_{in}^{p}}^{*}(k_{x_{1}},k_{y_{1}}) \label{Fpsi1}, \\
\widehat{\Psi}_{in}(
k_{x_{2}},k_{y_{2}})=\delta(k_{x_{2}},k_{y_{2}})+\sum_{q=1}^{\infty
}\frac{i^{q}}{q!}\widehat{\phi_{in}^{p}}(k_{x_{2}},k_{y_{2}})
\label{Fpsi2},
\end{align}
\end{subequations}

\noindent where $\delta(k_x,k_y)$ denotes the Dirac delta.

We may also write:

\begin{subequations}
\begin{align}
\overline{T}^{*}\left ( k_{x_{1}},k_{y_{1}}\right)=1+i^{*}\sum_{m,n=0}^{\infty }\alpha^{*}_{mn}k_{x_{1}}^{m}k_{y_{1}}^{n}, \label{Transf3} \\
\overline{T}\left ( k_{x_{2}},k_{y_{2}}\right )=1+i\sum_{\gamma,\nu =0}^{\infty }\alpha_{\gamma \nu}k_{x_{2}}^{\gamma}k_{y_{2}}^{\nu}. \label{Transf4}
\end{align}
\end{subequations}

We now substitute Eqs.~\eqref{Fpsi1}, \eqref{Fpsi2}, \eqref{Transf3} and \eqref{Transf3} into Eq.~\eqref{Cross2} and expand. A total of sixteen terms appear in the expansion making it a very lengthy expression to display, however similar mathematical manipulation is performed in each term which can be illustrated by using one term as an example. The longest term that appears is

\begin{eqnarray} \label{IntExp2}
\Bigg\langle \sum_{p,q,m,n,\gamma ,\nu =1}^{\infty }\frac{(i^{p+1})^{*}i^{q+1}\alpha^{*}_{mn}\alpha_{\gamma \nu}}{p!q!(i^{m+n})^{*}(i^{\gamma+\nu})} \nonumber\\
\frac{1}{2\pi}\iint_{-\infty}^{\infty}dk_{x_{1}}dk_{y_{1}}(i^{*}k_{x_{1}})^{m}(i^{*}k_{y_{1}})^{n}e^{[i^{*}(k_{x_{1}}x_{1}+k_{y_{1}}y_{1})]}  \nonumber\\
\times \widehat{\phi_{in}^{p}}^{*}(k_{x_{1}},k_{y_{1}}) \nonumber\\
\frac{1}{2\pi}\iint_{-\infty}^{\infty}dk_{x_{2}}dk_{y_{2}}(ik_{x_{2}})^{\gamma}(ik_{y_{2}})^{\nu}e^{[i(k_{x_{2}}x_{2}+k_{y_{2}}y_{2})]} \nonumber\\
\times \widehat{\phi_{in}^{q}}(k_{x_{2}},k_{y_{2}})\Bigg\rangle_{\omega}. \nonumber\\
\end{eqnarray}

By the Fourier derivative theorem \cite{Pag03}, the terms $(i^{*}k_{x_{1}})^{m}(i^{*}k_{y_{1}})^{n}e^{[i^{*}(k_{x_{1}}x_{1}+k_{y_{1}}y_{1})]}$ and $(ik_{x_{2}})^{\gamma}(ik_{y_{2}})^{\nu}e^{[i(k_{x_{2}}x_{2}+k_{y_{2}}y_{2})]}$ inside the double integrals can be expressed as $\partial^{m}_{x_{1}}\partial^{n}_{y_{1}}e^{[i^{*}(k_{x_{1}}x_{1}+k_{y_{1}}y_{1})]}$ and $\partial^{\gamma}_{x_{2}}\partial^{\nu} _{y_{2}}e^{[i(k_{x_{2}}x_{2}+k_{y_{2}}y_{2})]}$ respectively. The symbols $\partial^{m}_{x_{1}}$, $\partial^{n}_{y_{1}}$, $\partial^{\gamma}_{x_{2}}$ and $\partial^{\nu} _{y_{2}}$ denote partial derivatives with respect to the variables $x_{1}$, $y_{1}$, $x_{2}$ and $y_{2}$ and the indices ${m}$, ${n}$, ${\gamma}$ and ${\nu}$ are positive integers that denote the order of differentiation. With this re--expression we see that the integrals represent the inverse Fourier transforms of the functions $\partial^{m}_{x_{1}}\partial^{n}_{y_{1}}\widehat{\phi_{in}^{p}}(k_{x_{1}},k_{y_{1}})$ and $\partial^{\gamma}_{x_{2}}\partial^{\nu} _{y_{2}}\widehat{\phi_{in}^{q}}(k_{x_{2}},k_{y_{2}})$ and hence the entire term in Eq.~\eqref{IntExp2} simplifies to

\begin{eqnarray}     \label{TermExp01}
&&\sum_{p,q,m,n,\gamma ,\nu =1}^{\infty }\frac{(i^{p+1})^{*}i^{q+1}\alpha^{*}_{mn}\alpha_{\gamma \nu}}{p!q!(i^{m+n})^{*}(i^{\gamma+\nu})} \nonumber\\
&& \times \left \langle [\partial^{m}_{x_{1}}\partial^{n} _{y_{1}}\phi_{in}^{p}(x_{1},y_{1})]  [\partial^{\gamma}_{x_{2}}\partial^{\nu} _{y_{2}}\phi_{in}^{q}(x_{2},y_{2})]  \right \rangle_{\omega}. \nonumber\\
\end{eqnarray}

Using similar mathematical manipulation and logic used to get from Eq. \eqref{IntExp2} to \eqref{TermExp01} and applying it to all the terms which appear in the expansion of Eq. \eqref{Cross2}, one finds that the cross--spectral density for a shift--invariant linear system whose transfer function has infinitely many orders of aberrations is:

\begin{eqnarray}     \label{TermExp1}
W_{out}&=&1+\sum_{p=1}^{\infty }\frac{(i^{p})^{*}}{p!} \left \langle  \phi_{in}^{p}(x_{1},y_{1}) \right \rangle_{\omega} +\sum_{q=1}^{\infty }\frac{i^{q}}{q!} \left \langle  \phi_{in}^{q}(x_{2},y_{2}) \right \rangle_{\omega} \nonumber\\
& &+\sum_{p,q=1}^{\infty }\frac{(i^{p})^{*}i^{q}}{p!q!} \left \langle \phi_{in}^{p}(x_{1},y_{1}) \phi_{in}^{q}(x_{2},y_{2}) \right \rangle_{\omega} \nonumber\\
& &+ \sum_{p,m,n=1}^{\infty }
\frac{\alpha^{*}_{mn} (i^{p+1})^{*}}{p!(i^{m+n})^{*}}
 \left \langle \partial^{m}_{x_{1}}\partial^{n} _{y_{1}}\phi_{in}^{p}(x_{1},y_{1}) \right \rangle_{\omega} \nonumber\\
& & +\sum_{q,\gamma, \nu=1}^{\infty }
\frac{\alpha_{\gamma \nu} (i^{q+1})}{q!(i^{\gamma +\nu})} \nonumber\\
& & \times \left \langle \partial^{\gamma}_{x_{2}}\partial^{\nu} _{y_{2}}\phi_{in}^{q}(x_{2},y_{2}) \right \rangle_{\omega} \nonumber\\
& & +\sum_{p,q,m,n=1}^{\infty }\frac{\alpha^{*}_{mn}(i^{p+1})^{*}i^{q}}{p!q!(i^{m+n})^{*}}  \nonumber\\
& & \times \left \langle [\partial^{m}_{x_{1}}\partial^{n} _{y_{1}}\phi_{in}^{p}(x_{1},y_{1})] \phi_{in}^{q}(x_{2},y_{2}) \right \rangle_{\omega} \nonumber\\
& &+\sum_{p,q,\gamma,\nu=1}^{\infty }\frac{\alpha_{\gamma \nu}(i^{p+1})(i^{q})^{*}}{p!q!(i^{\gamma +\nu})} \nonumber\\
& & \times \left \langle \phi_{in}^{p}(x_{1},y_{1})[\partial^{\gamma}_{x_{2}}\partial^{\nu} _{y_{2}}\phi_{in}^{q}(x_{2},y_{2})]  \right \rangle_{\omega}  \nonumber\\
& &+\sum_{p,q,m,n,\gamma,\nu=1}^{\infty }\frac{(i^{p+1})^{*}i^{q+1}\alpha^{*}_{mn}\alpha_{\gamma \nu}}{p!q!(i^{m+n})^{*}(i^{\gamma +\nu})}  \nonumber\\
& & \times \left \langle [\partial^{m}_{x_{1}}\partial^{n} _{y_{1}}\phi_{in}^{p}(x_{1},y_{1})]  [\partial^{\gamma}_{x_{2}}\partial^{\nu} _{y_{2}}\phi_{in}^{q}(x_{2},y_{2})]  \right \rangle_{\omega}. \nonumber\\
\end{eqnarray}

This is a key result of the present paper.  We speak of it as exhibiting a $generalized$ $form$ $of$ $differential$ $phase$ $contrast$, in the sense that it is a representation in which the transverse derivatives of all orders of the phase distribution of each monochromatic field in the statistical ensemble, which are statistically averaged and weighted in constructing the output cross--spectral density.  The weighting coefficients are proportional to the generalized aberration coefficients drawn from the complex set $\left \{ \alpha_{mn} \right \}$, thereby demonstrating how individual generalized aberration coefficients contribute to particular orders of transverse derivative, of the phase of each monochromatic component in the statistical ensemble.


\section{Shift--invariant, linear systems for partially coherent fields considering different types of samples using one transverse spatial coordinate} \label{1Dsect}

In this section we will apply the formalism of the preceding section to three different types of sample, starting with samples which satisfy the weak--phase approximation (Sec.~\ref{WeakPsec}), followed by samples that satisfy the weak phase--amplitude approximation (Sec.~\ref{sec:WeakPhaseAmp}) and finally samples made from a single--material which also satisfy the weak phase--amplitude approximation (Sec.~\ref{sec:WeakSingleM}). In the interests of physical transparency of the resulting expressions, we will drop the number of transverse dimensions from two down to one.

\subsection{Samples that satisfy the weak--phase approximation} \label{WeakPsec}

The weak--phase approximation implies that when an object is illuminated by a wave--field the object itself causes very small changes in the phase of the incident field, as the scattering effects are relatively weak. We saw how under the ``phase object approximation" one may express $\Psi_{in}$ as a Taylor series (see Eq.\eqref{PsiTay2D}). In the one--dimensional perfectly coherent case this is written as

\begin{eqnarray} \label{PsiTay1D}
\Psi_{in}(x)=1+\sum_{p=1}^{\infty }\frac{i^{p}}{p!}\phi_{in}^{p}(x)
\end{eqnarray}

For samples which satisfy the weak--phase approximation, we can ignore anything higher than first--order terms in the phase, that is:

\begin{eqnarray}
\Psi_{in}(x)\approx 1+i\phi_{in}(x).
\end{eqnarray}

Physically, this corresponds to each strictly monochromatic component of the input statistical ensemble having a transverse phase variation whose magnitude is much smaller than one radian.  Such a strong limiting assumption of course implies significant loss of generality, a drawback which may be counterpointed with the very widespread use of the weak phase object approximation in visible--light imaging, x--ray imaging and electron imaging \cite{Cow01,Cleo01}.

In most cases relating to weak phase objects including such terms only up to first order in $\phi_{in}(x)$ is acceptable, however when we calculate the cross--spectral density $W_{out}$ we take the product of two wave fields which cause second--order terms in $\phi$ to appear which cannot be ignored. This simply means expanding sums over $p$ and $q$ in Eq. \eqref{TermExp1} up until terms that are no higher than second order in $\phi$. By doing this the one--dimensional version of the cross--spectral density $W_{out}$ for samples that satisfy the weak--phase approximation is

\begin{eqnarray} \label{Cross1D:A}
W_{out}&=&1+i^{*}\left \langle \phi_{in}(x_{1}) \right \rangle_{\omega}+i\left \langle \phi_{in}(x_{2}) \right \rangle_{\omega} \nonumber\\ 
& & -\frac{1}{2}\left \langle \phi_{in}^{2}(x_{1}) \right \rangle_{\omega}-\frac{1}{2}\left \langle \phi_{in}^{2}(x_{2}) \right \rangle_{\omega} \nonumber\\ 
& & +\left \langle \phi_{in}(x_{1}) \phi_{in}(x_{2}) \right \rangle_{\omega} \nonumber\\
& & -\sum_{m=1}^{\infty }\left ( \frac{\alpha_m}{i^m} \right )^{*}\left \langle  \partial^{m}_{x_{1}} \phi_{in}(x_{1})\right \rangle_{\omega} \nonumber\\
& &-\sum_{n=1}^{\infty }\left ( \frac{\alpha_n}{i^n} \right )\left \langle  \partial^{n}_{x_{2}} \phi_{in}(x_{2})\right \rangle_{\omega} \nonumber\\
& & -\sum_{m=1}^{\infty }\left ( \frac{\alpha_m}{i^{m-1}} \right )^{*}\left \langle  \partial^{m}_{x_{1}} \phi^{2}_{in}(x_{1})\right \rangle_{\omega} \nonumber\\
& &-\sum_{n=1}^{\infty }\left ( \frac{\alpha_n}{i^{n-1}} \right )\left \langle  \partial^{n}_{x_{2}} \phi^{2}_{in}(x_{2})\right \rangle_{\omega} \nonumber\\
& & +\sum_{m=1}^{\infty }\left ( \frac{\alpha_m}{i^{m-1}} \right )^{*}\left \langle  [\partial^{m}_{x_{1}} \phi_{in}(x_{1})]\phi_{in}(x_{2})\right \rangle_{\omega} \nonumber\\
& &+\sum_{n=1}^{\infty }\left ( \frac{\alpha_n}{i^{n-1}} \right )\left \langle \phi_{in}(x_{1}) [\partial^{n}_{x_{2}} \phi_{in}(x_{2})]\right \rangle_{\omega}  \nonumber\\
& & +\sum_{m,n=1}^{\infty }\left ( \frac{\alpha_m}{i^m} \right )^{*}\left ( \frac{\alpha_n}{i^n} \right ) \nonumber\\
& & \times \left \langle  [\partial^{m}_{x_{1}} \phi_{in}(x_{1})][\partial^{n}_{x_{2}} \phi_{in}(x_{2})]\right \rangle_{\omega}. \nonumber\\
\end{eqnarray}

The associated spectral density $S_{out}(x,\omega)\equiv W_{out}(x,x,\omega)$ is 

\begin{eqnarray} \label{Int1D:A}
S_{out}&=&1-2\sum_{m=1}^{\infty }{\textrm Re} \left (\frac{\alpha_m}{i^m} \right )\left \langle \partial^{m}_{x} \phi_{in}(x)\right \rangle_{\omega}   \nonumber\\
& &-2\sum_{m=1}^{\infty }{\textrm Re} \left (\frac{\alpha_m}{i^{m-1}} \right )\left \langle \partial^{m}_{x} \phi^{2}_{in}(x)\right \rangle_{\omega}  \nonumber\\
& &+2\sum_{m=1}^{\infty }{\textrm Re} \left (\frac{\alpha_m}{i^{m-1}} \right )\left \langle [\partial^{m}_{x} \phi_{in}(x)]\phi_{in}(x)\right \rangle_{\omega} \nonumber\\
& &+\sum_{m,n=1}^{\infty }\left (\frac{\alpha_m}{i^m} \right )^{*}\left (\frac{\alpha_n}{i^n} \right ) \nonumber\\
& & \times \left \langle [\partial^{m}_{x} \phi_{in}(x)][\partial^{n}_{x} \phi_{in}(x)]\right \rangle_{\omega}. \nonumber\\
\end{eqnarray}

Our earlier comments regarding generalized phase contrast are also applicable here.  Thus, for the case of weak phase objects imaged by an aberrated linear shift--invariant optical system, the output spectral density consists of a weighted sum of various orders of transverse derivative of the phases of each component of each strictly monochromatic member of the statistical ensemble quantifying the input stochastic process. The associated weighting coefficients are again proportional to the real or imaginary parts of the generalized aberration coefficients given by the complex set $\left \{ \alpha_{mn} \right \}$.

If we ignore terms in Eq.~\eqref{Int1D:A} that are higher than first order in $\phi$ and assume a perfectly coherent field (i.e. no ensemble average is required) then this equation reduces to the one dimensional form of the expression derived in the paper by Paganin and Gureyev \cite{Pag01} for linear shift--invariant imaging systems for fully coherent fields given by:

\begin{eqnarray} \label{PaganinGurEq13}
S_{out}=1-2\sum_{m=1}^{\infty }{\textrm Re} \left
(\frac{\alpha_m}{i^m} \right )\partial^{m}_{x} \phi_{in}(x).
\end{eqnarray}

Some interesting effects result when terms higher than first order in $\phi_{in}(x)$ are retained. For example, if we truncate Eq.~\eqref{Int1D:A} up to $m=1$ and $n=1$ the spectral density becomes

\begin{eqnarray} \label{Int1stOd:a}
S_{out}&=&1-2{\textrm Re}  \left (\frac{\alpha_{1}}{i} \right )\left \langle \partial_{x} \phi_{in}(x)\right \rangle_{\omega}-2{\textrm Re}(\alpha_1)\left \langle \partial_{x}\phi^{2}_{in}(x)\right \rangle_{\omega} \nonumber\\
& & +2{\textrm Re}(\alpha_1)\left \langle [\partial_{x} \phi_{in}(x)]\phi_{in}(x)\right \rangle_{\omega} \nonumber\\
& &+\left |\alpha_1 \right|^{2} \left \langle [\partial_{x} \phi_{in}(x)][\partial_{x} \phi_{in}(x)]\right \rangle_{\omega}. \nonumber\\
\end{eqnarray}

Here we have explicitly chosen a system that only displays first derivative contrast in the phase $\phi$. Now notice how invoking the product rule one may rewrite certain terms such as $\partial_{x}\phi^{2}_{in}=2[\partial_{x}\phi_{in}]\phi_{in}$ and $[\partial_{x}\phi_{in}][\partial_{x}\phi_{in}]=\partial_{x}([\partial_{x}\phi_{in}]\phi_{in})-[\partial^{2}_{x}\phi]\phi$ which makes Eq. \eqref{Int1stOd:a} appear as

\begin{eqnarray} \label{Int1stOd:b}
S_{out}&=&1-2{\textrm Re} \left (\frac{\alpha_{1}}{i} \right )\left \langle \partial_{x} \phi_{in}(x)\right \rangle_{\omega} \nonumber\\
& &-2{\textrm Re}(\alpha_1)\left \langle [\partial_{x} \phi_{in}(x)]\phi_{in}(x)\right \rangle_{\omega}  \nonumber\\
& & +\left |\alpha_1 \right |^{2} \left \langle \partial_{x}\left \{ [\partial_{x}\phi_{in}(x)]\phi_{in}(x)\right  \} \right \rangle_{\omega} \nonumber\\
& & -\left |\alpha_1 \right |^{2} \left \langle [\partial^{2}_{x}\phi_{in}(x)]\phi_{in}(x)\right \rangle_{\omega}. \nonumber\\
\end{eqnarray}

Notice how the final term yields a second derivative in the ensemble of phases. This is popularly referred to in the imaging field as ``Laplacian contrast" \cite{Teague01}. It is surprising that even though the system in Eq.~\eqref{Int1stOd:b} has been restricted to tilt aberrations $\alpha_{1}$ of first order that Laplacian contrast still arises.

\subsection{Samples that satisfy the weak phase--amplitude approximation} \label{sec:WeakPhaseAmp}

The next class of samples considered are those which satisfy the weak phase--amplitude approximation. This approximation takes into consideration the variations in both amplitude and phase that the wave--field incurs as it travels though the sample. Again, since we are working under the space--frequency description of partial coherence, these statements apply to each strictly monochromatic component of the illuminating beam which is elastically scattered by the sample to yield the ensemble of monochromatic fields which is input into the shift invariant linear imaging system.

Bearing the above in mind, the weak phase--amplitude approximation corresponds to the sample's scattering and absorptive properties being weak in the sense of the first Born approximation. For samples that induce changes in both phase and amplitude the one dimensional wave--field exiting is expressed as \cite{Cow01}:

\begin{eqnarray} \label{PsiWeakAmpCow}
\Psi_{in}(x)\equiv\exp[i\phi_{in}(x)-\mu_{in}(x)].
\end{eqnarray}
 
The real function function $\mu_{in}(x)$ is related to the transverse variations in intensity and like $\phi_{in}(x)$ it is  also a real function. It again proves convenient to express exponential functions as a Taylor series. In this case $\Psi_{in}(x)$ is given by

\begin{eqnarray} \label{PsiTayWA1D}
\Psi_{in}(x)=1+\sum_{p=1}^{\infty }\frac{[i\phi_{in}(x)-\mu_{in}(x)]^{p}}{p!}.
\end{eqnarray}

Like the weak--phase object approximation the weak phase--amplitude approximation also involves ignoring higher than first order terms allowing the wave--field to be expressed as

\begin{eqnarray} \label{PsiWeakAmp}
 \Psi_{in}(x)\approx 1+i\phi_{in}(x)-\mu_{in}(x).
\end{eqnarray}

Now, to obtain $W_{out}$ one simply needs to replace the terms $\phi_{in}(x_{1})$ and $\phi_{in}(x_{2})$ with $i\phi_{in}(x_{1})-\mu_{in}(x_{1})$ and $i\phi_{in}(x_{2})-\mu_{in}(x_{2})$ in Eq.~\eqref{Cross1D:A}, respectively. Note that second--order terms need to be included for the same reasons argued for the weak--phase approximation. Once we have $W_{out}$ then set $x_{1}=x_{2}=x$ to obtain an expression for the spectral density $S_{out}(x,\omega)$ for samples that are weak in phase and amplitude variations. In this case the spectral density is given by the following expression which again demonstrates generalized differential phase contrast in the sense defined earlier:

\begin{eqnarray} \label{IntWeakP,A}
S_{out}&=&1-2\left \langle \mu_{in}(x)\right \rangle_{\omega} 
-2\sum_{m=1}^{\infty }{\textrm Re} \left (\frac{\alpha_m}{i^m} \right )\left \langle \partial^{m}_{x} \phi_{in}(x)\right \rangle_{\omega} \nonumber\\
& &-2\sum_{m=1}^{\infty }{\textrm Re} \left (\frac{\alpha_m}{i^{m-1}} \right )\left \langle \partial^{m}_{x} \mu_{in}(x)\right \rangle_{\omega} \nonumber\\
& &-2\sum_{m=1}^{\infty }{\textrm Re} \left (\frac{\alpha_m}{i^{m-1}} \right )\left  \langle \partial^{m}_{x} \phi^{2}_{in}(x)\right \rangle_{\omega} \nonumber\\
& &+2\sum_{m=1}^{\infty }{\textrm Re} \left (\frac{\alpha_m}{i^{m-1}} \right )\left \langle \partial^{m}_{x} \mu^{2}_{in}(x)\right \rangle_{\omega} \nonumber\\
& &+6\sum_{m=1}^{\infty }{\textrm Re} \left (\frac{\alpha_m}{i^{m-2}} \right )\left \langle [\partial^{m}_{x} \phi_{in}(x)]\mu_{in}(x)\right \rangle_{\omega} \nonumber\\
& &+2\sum_{m=1}^{\infty }{\textrm Re} \left (\frac{\alpha_m}{i^{m-1}} \right )\left \langle [\partial^{m}_{x} \phi_{in}(x)]\phi_{in}(x)\right \rangle_{\omega} \nonumber\\
& &+2\sum_{m=1}^{\infty }{\textrm Re} \left (\frac{\alpha_m}{i^{m-2}} \right )\left \langle [\partial^{m}_{x} \mu_{in}(x)]\phi_{in}(x)\right \rangle_{\omega} \nonumber\\
& &+2\sum_{m=1}^{\infty }{\textrm Re} \left (\frac{\alpha_m}{i^{m-1}} \right )\left \langle [\partial^{m}_{x} \mu_{in}(x)]\mu_{in}(x)\right \rangle_{\omega} \nonumber\\
& &+\sum_{m,n=1}^{\infty }\left (\frac{\alpha_m}{i^{m}} \right )^{*} \left (\frac{\alpha_n}{i^{n}} \right )\left \langle [\partial^{m}_{x} \phi_{in}(x)][\partial^{n}_{x} \phi_{in}(x)]\right \rangle_{\omega} \nonumber\\
& &+\sum_{m,n=1}^{\infty }\left (\frac{\alpha_m}{i^{m}} \right )^{*} \left (\frac{\alpha_n}{i^{n}} \right )\left \langle [\partial^{m}_{x} \mu_{in}(x)][\partial^{n}_{x} \mu_{in}(x)]\right \rangle_{\omega}. \nonumber\\
\end{eqnarray}

Similar to the previous case if second order terms in $\phi$ and $\mu$ are neglected and we remove the angular brackets assuming a fully coherent wave then Eq. \eqref{IntWeakP,A} reduces to the one derived in Paganin and Gureyev \cite{Pag01} when dealing with the weak phase--amplitude approximation, namely:

\begin{eqnarray} \label{WeakAmp3}
S_{out}&=&1-2\mu_{in}(x)-2\sum_{m=1}^{\infty }{\textrm Re} \left (\frac{\alpha_m}{i^m} \right )\partial^{m}_{x} \phi_{in}(x) \nonumber\\
& &-2\sum_{m=1}^{\infty }{\textrm Re} \left (\frac{\alpha_m}{i^{m-1}} \right )\partial^{m}_{x} \mu_{in}(x). \nonumber\\
\end{eqnarray}

\subsection{Single--material samples that satisfy the weak phase--amplitude approximation} \label{sec:WeakSingleM}

The final kind of sample that we consider is those that are comprised of a single material and also have the transverse phase and intensity variations of the wave--field being small as it travels though the sample. Utilizing terminology commonly used by the x--ray optics community, assume that the single--material sample has a constant complex refractive index \cite{Pag03}:

\begin{eqnarray}
n=1-\delta+i\beta .
\end{eqnarray}

If the projected thickness along the orientation of a particular direction of propagation for paraxial illumination is denoted as $T_{proj}(x)$, then the real numbers $\delta$ and $\beta$ are related to functions $\phi_{in}(x)$ and $\mu_{in}(x)$ via \cite{Pag03}: 

\begin{subequations}
\begin{align}
\phi_{in}(x)=-k\delta T_{proj}(x), \label{Phi_Proj_approx} \\
\mu_{in}(x)=\beta kT_{proj}(x). \label{mu_Proj_approx}
\end{align}
\end{subequations}

\noindent Here, $k$ is the radiation wavenumber corresponding to the wavelength $\lambda$. This permits us to write the input wave--field as 

\begin{eqnarray}
\Psi_{in}(x)=\exp[k(\beta-i\delta)T_{proj}(x)],
\end{eqnarray}

Under the single--material weak phase--amplitude object approximation $\Psi_{in}(x)$ is approximated as

\begin{eqnarray}
\Psi_{in}(x)&=\exp[k(\beta-i\delta)T_{proj}(x)] \nonumber\\
&\approx 1-k(\beta-i\delta)T_{proj}(x).
\end{eqnarray}

Here we see that the ``single--material weak phase--amplitude object approximation" is none other than the ``weak phase--amplitude object approximation" that uses that fact that when a weak object is made out of only one material the functions $\phi_{in}(x)$ and $\mu_{in}(x)$ become proportional to each other. Bearing this in mind, to obtain an expression for the spectral density $S_{out}(x)$ for systems that are linear and shift--invariant when the object under study satisfies the ``single--material weak phase--amplitude object approximation" all that is needed is to replace $\phi_{in}(x)$ and $\mu_{in}(x)$ in Eq.\eqref{IntWeakP,A} with $-k\delta T_{proj}(x)$ and $\beta kT_{proj}(x)$ respectively to yield

\begin{eqnarray} \label{IntSingleWeakP,A}
S_{out}&=&1-2\beta k\left \langle T_{proj}(x)\right \rangle_{\omega}  \nonumber\\
& &-2\sum_{m=1}^{\infty }{\textrm Re} \left [\frac{\alpha_m k(i\beta+\delta)}{i^m} \right ]\left \langle \partial^{m}_{x} T_{proj}(x)\right \rangle_{\omega} \nonumber\\
& &+2\sum_{m=1}^{\infty }{\textrm Re} \left [\frac{\alpha_m k(\delta+\beta)}{i^{m-1}} \right ]\left \langle \partial^{m}_{x} T^{2}_{proj}(x)\right \rangle_{\omega} \nonumber\\
& &+2\sum_{m=1}^{\infty }{\textrm Re} \left [\frac{\alpha_m k^{2}(4\delta-\beta^{2}+i\delta^{2})}{i^{m}} \right ] \nonumber\\
& & \times \left \langle [\partial^{m}_{x}T_{proj}(x)]T_{proj}(x)\right \rangle_{\omega} \nonumber\\
& &+\sum_{m,n=1}^{\infty }\sigma\left (\frac{\alpha_m}{i^{m}} \right )^{*}\left (\frac{\alpha_n}{i^{n}} \right ) \nonumber\\ 
& & \times \left\langle [\partial^{m}_{x}T_{proj}(x)][\partial^{n}_{x}T_{proj}(x)]\right \rangle_{\omega}, \nonumber\\
\end{eqnarray}

\noindent where $\sigma=k^{2}(\delta+\beta)$, and ensemble averages are take over the sample projected thickness (i.e. $\left \langle T_{proj}(x)\right \rangle_{\omega}$). This implies taking the average sum of projected path integrals along the sample over a range of angular orientations, for the case where the incident ensemble of monochromatic fields consist of a set of plane waves. We see that the single--material assumption significantly simplifies the expression for the spectral density $S_{out}$. One of the advantages about making the ``single--material weak phase--amplitude object approximation" is that it allows one to relate the measured image directly to morphological detail of the sample bypassing the idea of ensembles of phase maps. For instance take a special case of Eq.~\eqref{IntSingleWeakP,A} where the system has a finite set of non--vanishing aberrations (all of which are known `a priori'), and we take the spectral density $S_{out}$ to be the measured quantity, leaving $\left \langle T_{proj}(x)\right \rangle_{\omega}$ as the unknown variable. This effectively brings about an inverse problem, where from an aberrated image one seeks to infer information about the size of sample. This is very common in the imaging world and can be related to the technique known as ``phase retrieval'', which as the name says involves retrieving the phase $\phi$ of the wave--field $\Psi$ once the field has travelled through the sample from either one or multiple intensity measurements. Usually this is done using some iterative or non--iterative algorithm and in most cases a perfectly coherent monochromatic wave--field is assumed. In the context of this paper we see that the idea of ``phase retrieval" is somewhat redundant since we have considered wave--fields that are partially coherent and that therefore do not have a characteristic phase $\phi$ but rather have a statistical signature $\left \langle \partial^{m}_{x} \phi_{in}\right \rangle_{\omega}$. This, highlights the importance of Eq.~\eqref{IntSingleWeakP,A} as it makes more sense to want to recover information about the morphology of the imaged sample as opposed to phase $\phi$ of a wave--field technically the latter does not exist in the context of partial coherence \cite{Wolf03}.


\section{The transfer function for shift--invariant linear systems with infinitely many orders of aberrations} \label{TransSec}

The transfer function formalism to study image formation is widely used to describe optical systems. This section discusses in detail the properties and characteristics of the transfer function used in the development of this theory (see Eq.~\eqref{Transf1}). The expressions for spectral densities for all three types of sample are derived under the Taylor series form of the transfer function which is written in terms of the coefficients $\alpha_{m}$. For this reason, it is important to state that actual aberration coefficients, namely those directly corresponding to the seidel aberrations, are those denoted by $\widetilde{\alpha}_{m}$. For example, $\widetilde{\alpha}_{2}$ is directly proportional to defocus ``$z$'' as is $\widetilde{\alpha}_{4}$ to spherical aberration \cite{Pag01} ``$C_{s}$''.  The main goal of this section will be to illustrate to the reader how we are able to express the transfer function as a Taylor--series expansion, which, eventually will lead us to another problem in finding a standard formula on how to relate the coefficients $\alpha_{m}$ to the aberration coefficients $\widetilde{\alpha}_{m}$, a problem which is solved using a combinatorial approach. Also, we will continue to use only one spatial dimension as in Sec.~\ref{1Dsect} in order to keep all mathematical manipulations simple. 

We begin by re--stating the transfer function in one spatial dimension

\begin{eqnarray}\label{Trans1D:1}
\overline{T}(k_{x})=\exp\left (i\sum_{m=0}^{\infty }\widetilde{\alpha}_{m}k_{x}^{m}\right )
\end{eqnarray}

We remind the reader that the set of complex numbers $\left \{\widetilde{\alpha}_{m} \right \}$ are labelled here as ``aberration coefficients'' whose object is to characterise a particular state of the linear imaging system. Each such coefficient is denoted

\begin{eqnarray}\label{alpha_complex}
\widetilde{\alpha}_{m}\equiv\widetilde{\alpha}^{(R)}_{m}+i\widetilde{\alpha}^{(I)}_{m},
\end{eqnarray}

\noindent where $\widetilde{\alpha}^{(R)}_{m}$ denotes the real part and $\widetilde{\alpha}^{(I)}_{m}$ denotes the imaginary part. It was assumed in Paganin and Gureyev \cite{Pag01} that at the Fourier--space origin the transfer function must equal unity, that is $\overline{T}(k_{x}=0)=1$. Such an assumption implies a trivial loss of generality for all systems that posses a transfer function that does not vanish at the Fourier space origin. Also under this assumption we may set $\widetilde{\alpha}_{0}=0$.

Now, we want to represent Eq.~\eqref{Trans1D:1} as a Taylor--series, something that in Paganin and Gureyev \cite{Pag01} was only stated but not shown. Here, we provide a more detailed explanation of how this is achieved. Firstly, let the entire sum in Eq. \eqref{Trans1D:1} be labelled $X \equiv i\sum_{m=1}^{\infty}\widetilde{\alpha}_{m}k_{x}^{m}$. The Taylor--series of an exponential function is given by

\begin{eqnarray}
e^{X}=1+\sum_{l=1}^{\infty}\frac{X^{l}}{l!},
\end{eqnarray}

\noindent where, $l$ is also a non--negative integer $l=1,2,...$. If we now substitute $X \equiv i\sum_{m=1}^{\infty}\widetilde{\alpha}_{m}k_{x}^{m}$ then Eq.~\eqref{Trans1D:1} becomes

\begin{eqnarray}
\overline{T}(k_{x})=1+i\sum_{l=1}^{\infty }\frac{i^{l-1}}{l!}\left (\sum_{m=1}^{\infty }\widetilde{\alpha}_{m}k_{x}^{m}\right )^{l}.
\end{eqnarray}

Notice how now we have commenced the summation from $m=1$. This is due to assumption made earlier that $\overline{T}(k_{x}=0)=1$ which in turn allowed to set $\widetilde{\alpha}_{0}=0$. Writing the summation $\sum_{m=1}^{\infty }\widetilde{\alpha}_{m}k_{x}^{m}$ explicitly we get

\begin{eqnarray} \label{SummationExample}
\overline{T}(k_{x})=1+i\sum_{l=1}^{\infty }\frac{i^{l-1}}{l!}\left (\widetilde{\alpha}_{1}k_{x}+\widetilde{\alpha}_{2}k^{2}_{x}+\widetilde{\alpha}_{3}k^{3}_{x}+\widetilde{\alpha}_{4}k^{4}_{x}+\cdots\right )^{l}\nonumber\\
\end{eqnarray}

We now turn our focus to the summation in Eq.~\eqref{SummationExample}. If one writes down the first few $l$ terms, say $l=1,2,3,4$, it can be seen that all the common powers of $k_{x}$ can be collected. For example:

\begin{eqnarray}
l=1,\: \: &\left (\widetilde{\alpha}_{1}k_{x}+\widetilde{\alpha}_{2}k^{2}_{x}+\widetilde{\alpha}_{3}k^{3}_{x}+\widetilde{\alpha}_{4}k^{4}_{x}+\cdots\right )^{1} \nonumber\\
l=2,\: \: &+\frac{i}{2!}\left (\widetilde{\alpha}_{1}k_{x}+\widetilde{\alpha}_{2}k^{2}_{x}+\widetilde{\alpha}_{3}k^{3}_{x}+\widetilde{\alpha}_{4}k^{4}_{x}+\cdots\right )^{2}\nonumber\\
l=3,\: \: &-\frac{1}{3!}\left (\widetilde{\alpha}_{1}k_{x}+\widetilde{\alpha}_{2}k^{2}_{x}+\widetilde{\alpha}_{3}k^{3}_{x}+\widetilde{\alpha}_{4}k^{4}_{x}+\cdots\right )^{3} \nonumber\\
l=4,\: \: &-\frac{-i}{4!}\left (\widetilde{\alpha}_{1}k_{x}+\widetilde{\alpha}_{2}k^{2}_{x}+\widetilde{\alpha}_{3}k^{3}_{x}+\widetilde{\alpha}_{4}k^{4}_{x}+\cdots\right )^{4}. \nonumber\\
\end{eqnarray}

Once we collect all the common powers of $k_{x}$ we see that the entire summation in Eq.~\eqref{SummationExample} can be expressed in the alternative form

\begin{eqnarray}\label{eqn:AlphaCollect}
&&\overset{\alpha_{1}}{\overbrace{(\widetilde{\alpha}_{1})}}k_{x}+\overset{\alpha_{2}}{\overbrace{(\widetilde{\alpha}_{2}+\frac{i}{2}\widetilde{\alpha}^{2}_{1})}}k^{2}_{x}+\overset{\alpha_{3}}{\overbrace{(\widetilde{\alpha}_{3}+i\widetilde{\alpha}_{1}\widetilde{\alpha}_{2}-\frac{1}{6}\widetilde{\alpha}^{3}_{1})}}k^{3}_{x}+ \nonumber\\
&&\overset{\alpha_{4}}{\overbrace{(\widetilde{\alpha}_{4}+i\widetilde{\alpha}_{1}\widetilde{\alpha}_{3}+\frac{i}{2}\widetilde{\alpha}^{2}_{2}-\frac{1}{2}\widetilde{\alpha}^{2}_{1}\widetilde{\alpha}_{2}-\frac{i}{24}\widetilde{\alpha}^{4}_{1})}}k^{4}_{x}+\cdots=\sum_{l=1}^{\infty }\alpha_{l}k_{x}^{l}. \nonumber\\
\end{eqnarray}

These mathematical manipulations reveal that we are able to represent the transfer function as the following Taylor--series.

\begin{eqnarray}
\overline{T}(k_{x})=1+i\sum_{m=1}^{\infty }\alpha_{m}k_{x}^{m}
\end{eqnarray}

We have re--labelled the non--negative integer $l$ with $m$ in order to remain consistent with our original notation. Also, notice that each $\alpha_{m}$ term is composed of a finite series of $\widetilde{\alpha}_{m}$ terms where the higher the order of $m$ the higher number the of terms that will appear. The fact that the series are finite turns out to be advantageous. On this note we see that another problem arises, that is, if one is dealing with aberrations that are higher in order than say $\alpha_{5}$, we saw from the above examples that computing all its terms in the series this can be tedious. This motivates us to seek a Standard Series Formula which can allow us to calculate any $\alpha_{m}$ for this problem by simply substituting fixed parameters to avoid such lengthy and tedious computations. This can be achieved if one visualises the problem as a combinatorial one. The first indication that tells us that this is solved combinatorially is when the term $\left (\widetilde{\alpha}_{1}k_{x}+\widetilde{\alpha}_{2}k^{2}_{x}+\widetilde{\alpha}_{3}k^{3}_{x}+\cdots\right )^{l}$ arises where we see that this is none other than a multinomial expansion which reveals its combinatorial nature. From this we can deduce that the terms in the series will have coefficients which can be calculated with the multinomial coefficients formula

\begin{eqnarray}
\binom{v}{m_{1},m_{2},\cdots ,m_{j}}=\frac{v!}{m_{1}!m_{2}!\cdots m_{j}!}.
\end{eqnarray}

Notice that for any $\alpha_{m}$ we find that the sum of the exponent times its subscript in each of its corresponding $\widetilde{\alpha}$ terms will always be equal. For instance take $\alpha_{3}=\widetilde{\alpha}_{3}+i\widetilde{\alpha}_{1}\widetilde{\alpha}_{2}-\frac{1}{6}\widetilde{\alpha}^{3}_{1}$; each of its $\widetilde{\alpha}$ terms in the expansion can be written as $\widetilde{\alpha}^{1}_{3}$, $\widetilde{\alpha}^{1}_{1}\widetilde{\alpha}^{1}_{2}$ and $\widetilde{\alpha}^{3}_{1}$. Now notice how the sum of the product of the exponents times its subscript for each term all equate to 3, we have $\widetilde{\alpha}^{1}_{3}$ ($1\times 3=3$), $\widetilde{\alpha}^{1}_{1}\widetilde{\alpha}^{1}_{2}$ ($1\times 1+1\times 2=3 $) and $\widetilde{\alpha}^{1}_{3}$ ($3\times 1=3$). If we do this for any $\alpha_{m}$ this condition will still hold.

Now our next step is to try to decode a particular pattern for any $\alpha_{m}$ series. Let's focus on $\alpha_{2}=\widetilde{\alpha}_{2}+\frac{i}{2}\widetilde{\alpha}^{2}_{1}$. Here we see that the highest power is $2$ and therefore one can also deduce that the highest power for any $\widetilde{\alpha}$ is never greater than $m$. We know that each expansion has a combinatorial nature so let's consider the terms that compose $\alpha_{2}$ are elements from the set $\left \{ \widetilde{\alpha}_{1},\widetilde{\alpha}_{2} \right \}$ and its corresponding exponents are combinations from the set $\left \{ 0,1,2 \right \}$. We also see that the coefficients will be given by the multinomial coefficients formula. If we write down all possible combinations with their corresponding coefficients it displays as

\begin{eqnarray}
&&\overset{0(1)+0(2)=0}{\frac{i^{-1}}{0!}\binom{0}{0,0}\widetilde{\alpha}^{0}_{1}\widetilde{\alpha}^{0}_{2}}+\overset{0(1)+1(2)=2}{\frac{i^{0}}{1!}\binom{1}{0,1}\widetilde{\alpha}^{0}_{1}\widetilde{\alpha}^{1}_{2}}+\overset{1(1)+0(2)=1}{\frac{i^{0}}{1!}\binom{1}{1,0}\widetilde{\alpha}^{1}_{1}\widetilde{\alpha}^{0}_{2}}\nonumber\\
&&+\overset{0(1)+2(2)=4}{\frac{i}{2!}\binom{2}{0,2}\widetilde{\alpha}^{0}_{1}\widetilde{\alpha}^{2}_{2}}+\overset{2(1)+0(2)=2}{\frac{i}{2!}\binom{2}{2,0}\widetilde{\alpha}^{2}_{1}\widetilde{\alpha}^{0}_{2}}+\overset{1(1)+1(2)=3}{\frac{i}{2!}\binom{2}{1,1}\widetilde{\alpha}^{1}_{1}\widetilde{\alpha}^{1}_{2}} \nonumber\\
\end{eqnarray}

As a convenient notation, notice that the sum of the product of the exponents times their corresponding subscripts have been deliberately placed above each combinatorial term. This helps us to see that if we only allow the terms in which the product of the exponents times their corresponding subscripts equals the order of the coefficient $\alpha_{m}$, in this case $m=2$ and neglect those which do not fulfill this condition then the surviving terms in the expansion will be the following 

\begin{eqnarray}
\alpha_{2}&=&\frac{i^{0}}{1!}\binom{1}{0,1}\widetilde{\alpha}^{0}_{1}\widetilde{\alpha}^{1}_{2}+\frac{i}{2!}\binom{2}{2,0}\widetilde{\alpha}^{2}_{1}\widetilde{\alpha}^{0}_{2} \nonumber\\
&=&\widetilde{\alpha}_{2}+\frac{i}{2}\widetilde{\alpha}^{2}_{1}.
\end{eqnarray}

Notice how applying this fusion of combinatorics and pattern decoding has arrived at the same answer for the $\alpha_{2}$ terms in Eq.~\eqref{eqn:AlphaCollect}. Now, we can employ the same strategy for $\alpha_{3}=\widetilde{\alpha}_{3}+i\widetilde{\alpha}_{1}\widetilde{\alpha}_{2}-\frac{1}{6}\widetilde{\alpha}^{3}_{1}$ where now all the terms are elements from the set $\left \{ \widetilde{\alpha}_{1},\widetilde{\alpha}_{2},\widetilde{\alpha}_{3} \right \}$ and its exponents are combinations from the set $\left \{ 0,1,2,3 \right \}$. Writing down the possible combinations will give

\begin{eqnarray}
&\overset{0(1)+0(2)+0(3)=0}{\frac{i^{-1}}{0!}\binom{0}{0,0,0}\widetilde{\alpha}^{0}_{1}\widetilde{\alpha}^{0}_{2}\widetilde{\alpha}^{0}_{3}}+\overset{1(1)+0(2)+0(3)=1}{\frac{i^{0}}{1!}\binom{1}{1,0,0}\widetilde{\alpha}^{1}_{1}\widetilde{\alpha}^{0}_{2}\widetilde{\alpha}^{0}_{3}}+\overset{0(1)+1(2)+0(3)=2}{\frac{i^{0}}{1!}\binom{1}{0,1,0}\widetilde{\alpha}^{0}_{1}\widetilde{\alpha}^{1}_{2}\widetilde{\alpha}^{0}_{3}}\nonumber\\
&+\overset{0(1)+0(2)+1(3)=3}{\frac{i^{0}}{1!}\binom{1}{0,0,1}\widetilde{\alpha}^{0}_{1}\widetilde{\alpha}^{0}_{2}\widetilde{\alpha}^{1}_{3}}+\overset{2(1)+0(2)+0(3)=2}{\frac{i^{1}}{2!}\binom{2}{2,0,0}\widetilde{\alpha}^{2}_{1}\widetilde{\alpha}^{0}_{2}\widetilde{\alpha}^{0}_{3}}+\overset{0(1)+2(2)+0(3)=4}{\frac{i^{1}}{2!}\binom{2}{0,2,0}\widetilde{\alpha}^{0}_{1}\widetilde{\alpha}^{2}_{2}\widetilde{\alpha}^{0}_{3}} \nonumber\\
&+\overset{0(1)+0(2)+2(3)=6}{\frac{i^{1}}{2!}\binom{2}{0,0,2}\widetilde{\alpha}^{0}_{1}\widetilde{\alpha}^{0}_{2}\widetilde{\alpha}^{2}_{3}}+\overset{1(1)+1(2)+0(3)=3}{\frac{i^{1}}{2!}\binom{2}{1,1,0}\widetilde{\alpha}^{1}_{1}\widetilde{\alpha}^{1}_{2}\widetilde{\alpha}^{0}_{3}}+\overset{1(1)+0(2)+1(3)=4}{\frac{i^{1}}{2!}\binom{2}{1,0,1}\widetilde{\alpha}^{1}_{1}\widetilde{\alpha}^{0}_{2}\widetilde{\alpha}^{1}_{3}} \nonumber\\
&+\overset{0(1)+1(2)+1(3)=5}{\frac{i^{1}}{2!}\binom{2}{0,1,1}\widetilde{\alpha}^{0}_{1}\widetilde{\alpha}^{1}_{2}\widetilde{\alpha}^{1}_{3}}+
\overset{3(1)+0(2)+0(3)=3}{\frac{i^{2}}{3!}\binom{3}{3,0,0}\widetilde{\alpha}^{3}_{1}\widetilde{\alpha}^{0}_{2}\widetilde{\alpha}^{0}_{3}}+\overset{0(1)+3(2)+0(3)=6}{\frac{i^{2}}{3!}\binom{3}{0,3,0}\widetilde{\alpha}^{0}_{1}\widetilde{\alpha}^{3}_{2}\widetilde{\alpha}^{0}_{3}}\nonumber\\
&+\overset{0(1)+0(2)+3(3)=9}{\frac{i^{2}}{3!}\binom{3}{0,0,3}\widetilde{\alpha}^{0}_{1}\widetilde{\alpha}^{0}_{2}\widetilde{\alpha}^{3}_{3}}+\overset{1(1)+1(2)+1(3)=6}{\frac{i^{2}}{3!}\binom{3}{1,1,1}\widetilde{\alpha}^{1}_{1}\widetilde{\alpha}^{1}_{2}\widetilde{\alpha}^{1}_{3}}+\overset{2(1)+1(2)+0(3)=4}{\frac{i^{2}}{3!}\binom{3}{2,1,0}\widetilde{\alpha}^{2}_{1}\widetilde{\alpha}^{1}_{2}\widetilde{\alpha}^{0}_{3}} \nonumber\\
&+\overset{2(1)+0(2)+1(3)=5}{\frac{i^{2}}{3!}\binom{3}{2,0,1}\widetilde{\alpha}^{2}_{1}\widetilde{\alpha}^{0}_{2}\widetilde{\alpha}^{1}_{3}}+\overset{1(1)+2(2)+0(3)=5}{\frac{i^{2}}{3!}\binom{3}{1,2,0}\widetilde{\alpha}^{1}_{1}\widetilde{\alpha}^{2}_{2}\widetilde{\alpha}^{0}_{3}}+\overset{1(1)+0(2)+2(3)=7}{\frac{i^{2}}{3!}\binom{3}{1,0,2}\widetilde{\alpha}^{1}_{1}\widetilde{\alpha}^{0}_{2}\widetilde{\alpha}^{2}_{3}}\nonumber\\
&+\overset{0(1)+1(2)+2(3)=8}{\frac{i^{2}}{3!}\binom{3}{0,1,2}\widetilde{\alpha}^{0}_{1}\widetilde{\alpha}^{1}_{2}\widetilde{\alpha}^{2}_{3}} \nonumber\\
\end{eqnarray}

Like the case for $\alpha_{2}$, if we only consider the terms where the sum of the exponents times their corresponding subscript equal $m=3$ and neglect the rest then the only terms which survive are

\begin{eqnarray}
\alpha_{3}&=&\overset{0(1)+0(2)+1(3)=3}{\frac{i^{0}}{1!}\binom{0}{0,0,1}\widetilde{\alpha}^{0}_{1}\widetilde{\alpha}^{0}_{2}\widetilde{\alpha}^{1}_{3}}+\overset{1(1)+1(2)+0(3)=3}{\frac{i^{1}}{2!}\binom{2}{1,1,0}\widetilde{\alpha}^{1}_{1}\widetilde{\alpha}^{1}_{2}\widetilde{\alpha}^{0}_{3}}\nonumber\\
&&+\overset{3(1)+0(2)+0(3)=3}{\frac{i^{2}}{3!}\binom{3}{3,0,0}\widetilde{\alpha}^{3}_{1}\widetilde{\alpha}^{0}_{2}\widetilde{\alpha}^{0}_{3}} \nonumber\\
&=&\widetilde{\alpha}_{3}+i\widetilde{\alpha}_{1}\widetilde{\alpha}_{2}-\frac{1}{6}\widetilde{\alpha}^{3}_{1} \nonumber\\
\end{eqnarray}

By extending the above logic one is able deduce the following standard formula to compute any $\alpha_{m}$:

\begin{eqnarray} \label{standardFormula}
\alpha_{m}=\sum_{v}^{m}\sum_{m_{1}+m_{2}+...+m_{j}=v}\frac{i^{v-1}}{v!}\binom{v}{m_{1},m_{2},\cdots,m_{j}}\nonumber\\
\widetilde{\alpha}^{m_{1}}_{1}\widetilde{\alpha}^{m_{2}}_{2}...\widetilde{\alpha}^{m_{j}}_{j} \nonumber\\
{\textrm w}{\textrm h}{\textrm e}{\textrm r}{\textrm e},\sum_{j}m_{j}\times j=m. \nonumber\\
\end{eqnarray}

Here, $v=0,1,2,...,m$, $j=1,2,...,m$ and $m_{j}=0,1,2,...,m$. To verify this standard formula, we calculate another $\alpha_{m}$ and see if we arrive at the same result to that obtained by collecting terms as done previously in Eq.~\eqref{eqn:AlphaCollect}. We do this by calculating $\alpha_{4}$ where according to the set condition one only needs to consider the terms which satisfy$\sum_{j}m_{j}\times j=4$. Below we display all the relevant terms

\begin{eqnarray}
\alpha_{4}&=&\overset{0(1)+0(2)+0(3)+1(4)=4}{\frac{i^{0}}{1!}\binom{1}{0,0,0,1}\widetilde{\alpha}^{0}_{1}\widetilde{\alpha}^{0}_{2}\widetilde{\alpha}^{0}_{3}\widetilde{\alpha}^{1}_{4}}+\overset{1(1)+0(2)+1(3)+0(4)=4}{\frac{i}{2!}\binom{2}{1,0,1,0}\widetilde{\alpha}^{1}_{1}\widetilde{\alpha}^{0}_{2}\widetilde{\alpha}^{1}_{3}\widetilde{\alpha}^{0}_{4}} \nonumber\\
&&+\overset{0(1)+2(2)+0(3)+0(4)=4}{\frac{i}{2!}\binom{2}{0,2,0,0}\widetilde{\alpha}^{0}_{1}\widetilde{\alpha}^{2}_{2}\widetilde{\alpha}^{0}_{3}\widetilde{\alpha}^{0}_{4}}+\overset{2(1)+1(2)+0(3)+0(4)=4}{\frac{i^{2}}{3!}\binom{3}{2,1,0,0}\widetilde{\alpha}^{2}_{1}\widetilde{\alpha}^{1}_{2}\widetilde{\alpha}^{0}_{3}\widetilde{\alpha}^{0}_{4}} \nonumber\\
&&+\overset{4(1)+0(2)+0(3)+0(4)=4}{\frac{i^{3}}{4!}\binom{4}{4,0,0,0}\widetilde{\alpha}^{4}_{1}\widetilde{\alpha}^{0}_{2}\widetilde{\alpha}^{0}_{3}\widetilde{\alpha}^{0}_{4}} \nonumber\\
&=&\widetilde{\alpha}_{4}+i\widetilde{\alpha}_{1}\widetilde{\alpha}_{3}+\frac{i}{2}\widetilde{\alpha}^{2}_{2}-\frac{1}{2}\widetilde{\alpha}^{2}_{1}\widetilde{\alpha}_{2}-\frac{i}{24}\widetilde{\alpha}^{4}_{1}
\end{eqnarray}

The computation above is in agreement with calculating $\alpha_{4}$ via the standard formula and the more lengthy method which involves collecting term of $k_{x}^{4}$ powers as was done in Eq.~\eqref{eqn:AlphaCollect}.
 
To end this section we return to the case where the transfer function contains two spatial dimensions transverse to the imaging direction. For such cases the combinatorial analysis is more complex. Nevertheless, using a similar strategy to the one used to formulate the standard formula for the one--dimensional case one is also able to deduce a formula for the more common imaging scenario with two spatial dimensions. Setting $\widetilde{\alpha}_{00}=0$ yield a ``two spatial dimensions standard formula'' of the form  

\begin{eqnarray} \label{standardFormula2D}
\alpha_{mn}=\sum_{v}^{m+n}\sum_{m_{01}+m_{10}+...+m_{j\nu}=v}\frac{i^{v-1}}{v!}\binom{v}{m_{01},m_{10},\cdots,m_{j\nu}}  \nonumber\\
\widetilde{\alpha}^{m_{01}}_{01}\widetilde{\alpha}^{m_{10}}_{10}...\widetilde{\alpha}^{m_{j\nu}}_{j\nu} \nonumber\\
{\textrm w}{\textrm h}{\textrm e}{\textrm r}{\textrm e},\sum_{j,\nu}m_{j\nu}\times (j+\nu)=m+n  \nonumber\\
\end{eqnarray}

\noindent where $v=0,1,2,...,m+n$, $j=0,1,2,...,m$ $\nu=0,1,2,...,n$ and $m_{j\nu}=0,1,2,...,m+n$. Since we have set $\widetilde{\alpha}_{00}=0$ we must impose the condition that when $j=0$ then $\nu \neq 0$ and vice versa. It is possible to make further simplification of the formula if rotational symmetry is also assumed (i.e. $\widetilde{\alpha}_{10}=\widetilde{\alpha}_{01}$).  

\section{Discussion and Summary}

In this work we have treated the problem of aberrations for partially coherent complex scalar wave--fields imaged by optical systems characterized by a transfer function which is both linear and shift--invariant. We have derived expressions for the output cross--spectral density $W_{out}$ using only one spatial variable, for members of the ensemble that satisfy the ``phase object approximation", the ``weak--phase object approximation", the ``weak phase--amplitude approximation" and finally the ``single--material weak phase--amplitude approximation", for certain classes of specimens. Also, for the three classes of  samples mentioned, an expression for the spectral density $S_{out}$ was calculated in which we saw how under certain restrictions the equations reduced to those derived in Paganin and Gureyev \cite{Pag01} where partial coherence is not considered. For the single material case the idea of ``phase retrieval'' was mentioned, however in the context of partial coherence this was rather redundant but nonetheless gave rise to the idea of carrying out morphological studies of imaged samples partially coherent light and aberrated imaging systems. This idea may have several applications in many areas such as geology, microbiology, material science, etc.

The transfer function was studied where we emphasized how one obtains the coefficients of Taylor--series representation of the transfer function.  This lead to a different problem involved finding a standard formula which can allow the calculation of any coefficient $\alpha_{m}$ in terms of its corresponding aberration coefficients $\widetilde{\alpha_{m}}$ for an infinite number of aberration orders. This standard formula brings many advantages not only in the sense that it is not limited to a finite  order of aberrations but also allows for broader considerations in ``aberration balancing". Aberration balancing is the act of seeking certain conditions in which the aberrations present in an optical system are negated by the system itself. To be more concise, one seeks to balance out the aberrations in an optical system against one another. This is somewhat similar to the notion of Scherzer defocus, where defocus is tuned to balance out spherical aberration \cite{Scher01}. For example, consider the spectral density in Eq.~\eqref{WeakAmp3} for the ``weak--phase object approximation" case. Suppose one aimed to find the conditions for which all aberrations present balanced out one another such that the output image displayed only first order differential contrast, that is, 

\begin{eqnarray} 
S_{out}=1-2\left \langle \partial_{x} \phi_{in}(x)\right \rangle_{\omega}.
\end{eqnarray}

This would require the following balancing conditions in order to achieve such an output image:

\begin{eqnarray}
& &\alpha_{1}^{(I)}=1, \overset{ m\neq 0}{{\textrm Re} \left ( \frac{\alpha_{m}}{i^{m}} \right )=0}, {\textrm Re} \left ( \frac{\alpha_{m}}{i^{m-1}} \right )=0,\nonumber\\
& & \left ( \frac{\alpha_{m}}{i^{m}} \right )^{*}\left( \frac{\alpha_{n}}{i^{n}} \right )=0. \nonumber\\
\end{eqnarray}

For systems with infinitely many aberrations (Eq.~\eqref{standardFormula}) the above balancing equation could be in principle solved with the help of the standard formula without the need for truncating the system.

\begin{acknowledgments}
M. A. Beltran acknowledges funding from the Monash University Dean's Scholarship Scheme. M. J. Kitchen acknowledge funding from the Australian Research Council (ARC, DP110101941).
\end{acknowledgments}

\end{document}